\documentclass[twocolumn,showpacs,preprintnumbers,amsmath,amssymb,prb]{revtex4}

\usepackage{graphicx}
\usepackage{dcolumn}
\usepackage{bm}

\begin{document}

\newpage

\title{Local oxidation of Ga[Al]As heterostructures with modulated tip-sample voltages}

\author{D. Graf,\renewcommand{\thefootnote}{\alph{footnote})}\footnote{Author to whom correspondence should be addressed; electronic mail:grafdavy@phys.ethz.ch} M. Frommenwiler, P. Studerus, T. Ihn, K. Ensslin}%
\affiliation{Solid State Physics Laboratory - ETH Z\"urich, Switzerland}
\author{D.C. Driscoll}%
\author{A.C. Gossard}
\affiliation{Materials Departement, University of California - Santa Barbara, Ca 93106, USA}%

\date{\today}

\begin{abstract}
Nanolithography based on local oxidation with a scanning force microscope has been performed on an undoped GaAs wafer and a Ga[Al]As heterostructure with an undoped GaAs cap layer and a shallow two-dimensional electron gas. The oxide growth and the resulting electronic properties of the patterned structures are compared for constant and modulated voltage applied to the conductive tip of the scanning force microscope. All the lithography has been performed in non-contact mode. Modulating the applied voltage enhances the aspect ratio of the oxide lines, which significantly strengthens the insulating properties of the lines on GaAs. In addition, the oxidation process is found to be more reliable and reproducible. Using this technique, a quantum point contact and a quantum wire have been defined and the electronic stability, the confinement potential and the electrical tunability are demonstrated to be similar to the oxidation with constant voltage.
\end{abstract}

\pacs{Valid PACS appear here}


\maketitle


\section{\label{sec:intro}Introduction}

Combining anodic oxidation and scanning probe microscopy, first with the scanning tunneling microscope (STM),\cite{Dagata90,Snow93} later with the scanning force microscope (SFM),\cite{Day93} has opened the way for patterning surfaces selectively in the submicron regime. 
Apart from using the oxide pattern as an etching mask or as a substrate for material growth, Ishii et al. demonstrated that an oxide line can induce a change in the resistance of a buried two-dimensional electrons gas (2DEG).\cite{Ishii95}  Held et al. showed that oxide lines deplete a shallow 2DEG in a Ga[Al]As heterostructures (less than 50 nm from the surface) completely,\cite{Held98} providing a versatile tool for the design of nanostructures with arbitrary shape and allowing topographical inspection without changing setup. \cite{Fuhrer02} The oxide height leading to a depletion of the electron gas must be chosen according to the distance of the 2DEG to the sample surface, which can not be too close without increasing significantly the scattering of the electrons at the doping atoms situated between 2DEG and surface. The deeper the 2DEG is located below the surface, the higher the height of the oxide line required for the realization of a lateral insulator. We are therefore interested in the regime, where the oxide height is 10 nm and beyond. 

Historically, one crucial aspect was a better understanding of the physical and chemical mechanisms dominating the oxidation process, in particular on doped silicon.\cite{Avouris97,Garcia98,Garcia99,Stievenard97} Two issues are closely related in view of optimal device fabrication: improving the lateral resolution and overcoming the self-limiting growth. They can be quantitatively described with the aspect ratio, which is defined as the ratio of the height and the base width of the oxide.
Dagata et al. pointed to the traps of ionic charges accumulating in the oxide and to the charge diffusion in the water meniscus forming between tip and sample as major contributions, referring to it as the space charge model.\cite{Dagata98} Pulsed or modulated (AC) voltage instead of constant (DC) voltage applied between the tip and the sample was suggested to neutralize the reaction products and limit the lateral carrier diffusion. Experimentally, enhanced growth in the central region of the water meniscus at the apex of the tip has been reported on n- and p-doped Si,\cite{Dagata98,Perez99,Calleja99} Ti,\cite{Dagata98} and n-GaAs,\cite{Okada00b} as well as p-GaAs.\cite{Matsuzaki01} The success on such a broad set of substrate materials suggests that the space charge model is applicable to local anodic oxidation in general and not restricted to a specific electro-chemical reaction, which is still debated in the case of GaAs.\cite{Okada00,Lazzarino05}

In this paper we investigate lines oxidized with constant and modulated voltages on undoped GaAs and discuss their influence on the electronic properties of a shallow 2DEG. All the lithography has been performed in non-contact mode.
In Sec.~\ref{sec:locox} we make a topographic comparison between DC and AC voltage oxidation exploring the rich parameter space. Our results are in qualitative agreement with the work of other groups and can be linked to established models for local anodic oxidation.
The analysis of a large set of oxidation trials showed two important results: 1) A clear improvement of the reliability and reproducibility when writing with AC voltage, which is encouraging in view of the design of more complex nanostructures, 2) an enhancement of the aspect ratio of about 15 \%, which, translated into electronic properties, represents a considerable gain in electronic depletion and scaling towards more dense device architectures (relevant quantum length scale is the Fermi wave length of about 40 nm). 
In Sec.~\ref{sec:algas} the two oxidation schemes are applied to deplete a 2DEG in a Ga[Al]As heterostructure. A quantitative characterization of the breakdown voltage versus oxide shape is presented, which turns out to be highly sensitive. We report on transport measurements on quantum point contacts and quantum wires written with DC as well as with AC voltage. Qualitatively the AC oxidation can compete with DC oxidation in terms of electronic stability, confinement potential and tunability.

\section{\label{sec:experiment} Experiment}

In local anodic oxidation a conductive SFM tip acts as the cathode whereas the grounded sample is the anode. A voltage is applied to the tip, while the sample is grounded via a metallic underlay. A thin water film covering the sample surface serves as the electrolyte. The ambient humidity is therefore actively controlled in our setup (42 \%). Both, for scanning and writing, the SFM (Asylum Research MFP-3D) is operated in non-contact mode with a resonance frequency between 315-325 kHz (force constant: 14-40 N/m). The full tip cone angle is between 20-30 degree, the tip curvature radius is less than 40 nm. The coating evaporated on n-doped Silicon tip corresponds to 20 nm Ti / 10 nm Pt or 25 nm W$_{2}$C. The feedback loop with amplitude damping as a control parameter is active during the writing process, which is crucial for writing more complex nanodevices consisting of many line sections. 

In the following we abbreviate 'oxidation with a constant/modulated voltage between tip and sample' with 'constant (DC)/modulated (AC) oxidation'. In the AC oxidation the tip bias consists of a square periodic signal [inset Fig.~\ref{fig:resettime}(a)]. Compared to DC oxidation, it offers more tuning parameters. Apart from the oxidation voltage $V_{ox}$, tip speed $v$ and set-point, which are also controlled during the DC oxidation, we can change the modulation frequency $f$, the reset voltage $V_{res}$ and the oxidation time $t_{ox}$. In all cases we used a moderate positive reset voltage between 2 and 4 V in order to ensure the trapped ionic charges to be neutralized, but avoiding oxidation of the tip.\cite{Perez99}

Taking the writing velocity of 0.2 $\mu$m/s and the base width of the oxide line of about 100 nm into account, a single spot on a line is not oxidized longer than 1 s. This is in contrast to previous experiments on n- and p-doped GaAs, where dots with oxidizing times of the order of several seconds were considered.\cite{Okada00b,Matsuzaki01} 
The resonance frequency of the cantilever is two orders of magnitude larger than the frequency of the AC voltage. There are two consequences: First, for each oxidation cycle ($t_{ox}+t_{res}$) we integrate over about $10^3$ cantilever oscillations; second, the feedback loop is able to control the tip-sample distance during the individual oxidation and reset stages. We thus operate in a low-frequency regime. Other experiments have coupled the resonance frequency of the cantilever to the modulation voltage making the phase difference an adjustable parameter for the oxidation performance.\cite{Legrand99}

\section{\label{sec:results} Results and discussion}

\subsection{\label{sec:locox}\normalsize G\scriptsize a\normalsize A\scriptsize s \normalsize wafer: Topographic properties}

As a first application of the technique, we show the oxidation of an undoped GaAs wafer (Liquid Encapsulated Czochralski). We precleaned the samples in an aceton/isopropanol/water ultrasonic bath sequence with additional plasma etching (O$_2$, 120 s, at a power of 200 W). No special wet chemical treatment was used to remove the native oxide. Since our motivation is the fabrication of devices on shallow 2DEGs, we focus on oxide lines with a given height of approximatly 10 nm, representing the basic building blocks for quantum point contacts, quantum wires and quantum dots. 

In Sec.~\ref{sec:freq} we will focus on the frequency dependence of the oxide height and the aspect ratio, whereas in Sec.~\ref{sec:other} other parameters will be discussed, such as oxidation voltage, duty ratio, set-point and tip velocity. In Sec.~\ref{sec:reli} we demonstrate the improved writing reliability based on a large collection of oxide lines.

\subsubsection{\label{sec:freq}Dependence on the frequency of the modulated voltage}

\begin{figure}
\includegraphics[width=0.4\textwidth]{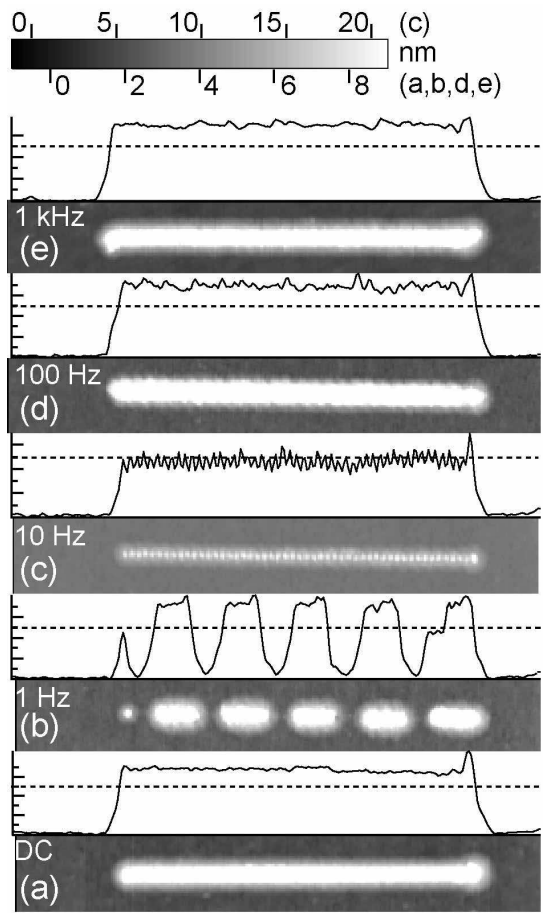}
\caption{\label{fig:linesfreq} SFM images (1.5 $\mu$m stripes, lines 1 $\mu$m long) with corresponding longitudinal oxide height profile (dashed line marks 10 nm in height) for (a) constant (DC) oxidation voltage and modulated (AC) voltages: (b) $f$=1 Hz, (c) $f$=10 Hz, (d) $f$=100 Hz and (e) $f$=1 kHz. The lines were written from right to left. For 1 Hz (b) and 10 Hz (c) the pulse scheme of the oxidation voltage is apparent in the oxide height profiles, whereas for higher frequencies (d),(e) the individual sections merge to a continuous line as for the DC case (a). Parameters used: $V_{ox}$=-16 V, $V_{res}$=4 V, $t_{res}$/($t_{res}$+$t_{ox}$)=50 \% (30 \% for $f$=1 kHz). Note that for (c) a different gray scale (top) is used.}
\end{figure}

In Fig.~\ref{fig:linesfreq} SFM micrographs and height profiles of oxide lines written with DC and AC voltages are presented. With a writing velocity of $v=0.2$ $\mu$m/s, the 1 Hz line is just an on/off cycling of shorter DC sections. In the 10 Hz case the individual dots can still be seen if the right contrast is chosen. For further increased frequencies up to 1 kHz consecutive pulses ($t_{ox}+t_{res}$) merge to form a continuous line. Capacitive losses leading to a low-pass filter effect limit the operational bandwidth to a few kHz. 

\begin{figure}
\includegraphics[width=0.4\textwidth]{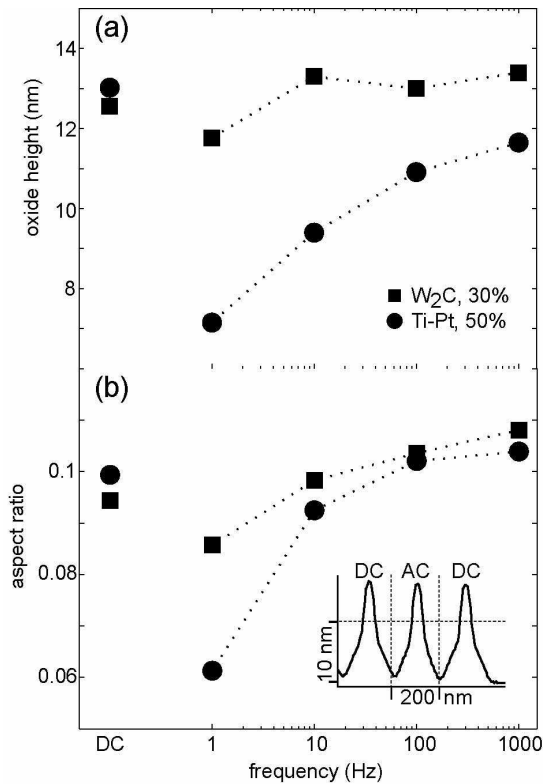}
\caption{\label{fig:frequency} Oxide height (average over the line) and aspect ratio for different modulation frequencies compared with DC writing. Although the total oxidation time is reduced by 30\% or 50 \% compared to the DC case, we reach practically the same oxide height, but a clearly better aspect ratio at 1 kHz. The tip velocity is fixed at 0.2 $\mu$m/s for all lines. Note that different SFM tips have been used (W$_{2}$C / Ti-Pt). Inset: Profile of neighbouring DC and AC lines. The total oxidation dose is kept constant by adjusting the writing speed: $v$(DC)=0.3 $\mu$m/s / $v$(AC)=0.15 $\mu$m/s, duty cycle = 50 \%.}
\end{figure}

Since for 1 and 10 Hz we get a clear segmentation of oxidized and non-oxidized regions [Fig.~\ref{fig:linesfreq}(b),(c)], it is not surprising that the average height taken over all the sections of the line, as shown in Fig.~\ref{fig:frequency}, is reduced for low modulation frequencies. However, we stress the fact that for higher frequencies the height of the AC voltage lines reaches almost the one of the constant voltage (DC) lines, despite the fact that the anodizing time has been reduced by 30 \% or 50 \% since we kept the writing speed constant.

The same behavior is observed in the aspect ratio, since the reduction for low frequencies is mostly due to the height averaging. We find that the aspect ratio for $f=1$ kHz is enhanced compared to the DC value. 
On the other hand we can keep the total oxidizing time constant by varying the writing speed. For a duty cycle of 50 \% this implies doubling the writing speed in the DC compared to the AC case. We find [inset Fig.~\ref{fig:frequency}(b)], that the line height does not vary, but a significant narrowing of the base width can be observed, indicating a reduced lateral carrier diffusion.\cite{Calleja00}
Corresponding values for a different set of lines are presented in Table~\ref{tab:aspect}. The enhancement of about 15 \% in the aspect ratio has a dramatic effect on the insulating properties of depleted lines in an underlying two-dimensional electrons gas, as discussed in Sec.~\ref{sec:algas}.

\begin{table}
\caption{\label{tab:aspect} The average and standard deviation of two sets of lines (\#DC=18 lines,\#AC=8 lines) with $V_{ox}$=-18 V fixed are listed (further AC parameters: $V_{res}$=2 V, $t_{res}$/($t_{res}$+$t_{ox}$)=30 \%, $f$=1 kHz). SFM tip used: Ti-Pt.}
\begin{ruledtabular}
\begin{tabular}{ccccc}
 & height (nm) & width (nm) & aspect ratio & FWHM (nm) \\
			\hline
			DC& $13.3 \pm 2.5$ & $133.0 \pm 12.6$ & $0.1002 \pm 0.0139$ & $53.6 \pm 8.0$ \\
			AC& $15.2\pm 0.5$ & $132.4\pm 3.2$ & $0.1150\pm 0.0052$ & $50.0\pm 3.1$ \\
			\hline
			 & +14.3 \% & -0.5 \% & +14.8 \% & -6.7 \% \\
\end{tabular}
\end{ruledtabular}
\end{table}

\subsubsection{\label{sec:other}Dependence on other parameters}

The height of the oxide lines increases linearly with the negative oxdiation voltage $V_{ox}$, in the case of constant voltage with approximately 1 nm per 1 V (data not shown). A similar relation is observed for modulated voltage as shown in Table~\ref{tab:offset}, but the oxidation with modulated voltage opens the door to a much larger parameter space. 

\begin{figure}
\includegraphics[width=0.4\textwidth]{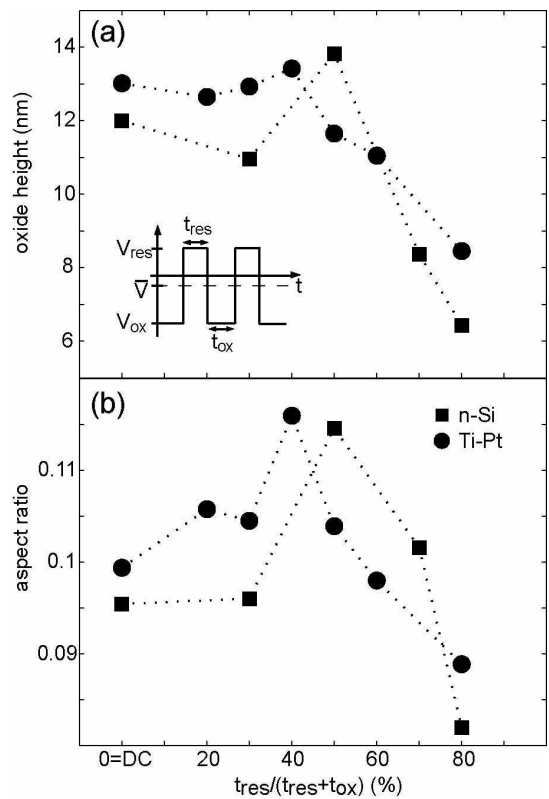}
\caption{\label{fig:resettime} Reset time in percents of the total pulse cycle. The oxide height (a) stays almost constant, but diminishes when the reset time exceeds the oxidation time. Compared to that we find a maximum in the aspect ratio (b) around 40 to 50 \%. The different symbols correspond to two different tip materials. Inset: Rectangular pulse scheme used for modulation voltage.}
\end{figure}

In addition to the frequency already discussed in the previous section, the duty ratio, i.e. the relation between reset and oxidation duration, can be chosen at will. The oxidation is due to its fast initial growth rate sensitive to time scales in the sub-millisecond regime.\cite{Avouris97} Thus changing the duty ratio will have an effect on the oxide height as a result of different oxidation times $t_{ox}$ [Fig.~\ref{fig:resettime}(a)]. For a duty ratio higher than 40 \% - 50 \% we find a linear dependence of the oxide height versus effective oxidation time. Lower duty ratios do not seem to affect the height. The aspect ratio has a non-monotonic behavior with respect to the duty ratio [Fig.~\ref{fig:resettime}(b)]. The aspect ratio shows a maximum along the duty cycle axis around 40 \% - 50 \%, which is clear evidence that a voltage modulation on these time scales influences the oxide growth. Note that two different tip materials have been used (n-Si / Ti-Pt): The results are in good qualitative agreement, even though they are horizontally shifted by about 10 \%. The same behavior has also been reported for p-doped GaAs.\cite{Matsuzaki01}  

\begin{table}
\caption{\label{tab:offset} Oxide height and aspect ratio for two different offset voltages. The set counts 27 ($\protect \bar{V}$=-6 V) respectively 8 ($\bar{V}$=-8 V) different AC lines. Further parameters are: $v$=200 $\mu$m/s, $t_{res}$/($t_{res}$+$t_{ox}$)=30 \%, $f$=1 kHz.  Note that since the peak-to-peak amplitude of the modulated voltage is unchanged, not only $V_{ox}$ but also $V_{res}$ is shifted parallel to the offset. SFM tip used: Ti-Pt. For definition of $\bar{V}$, $V_{ox}$ and $V_{res}$ see inset of Fig.~\protect \ref{fig:resettime}(a).}
\begin{ruledtabular}
\begin{tabular}{c c c c c}
			$\bar{V}$ (V) & $V_{ox}$ (V) & $V_{res}$ (V) & oxide height (nm) & aspect ratio\\
			\hline
			-6 & -16 & 4 & $12.9\pm1.0$ & $0.1045\pm0.0085$ \\
			-8 & -18 & 2 & $15.2\pm0.5$ & $0.1150\pm0.0052$ \\
\end{tabular}
\end{ruledtabular}
\end{table}

It has been stated in the literature that non-contact mode is more gentle to the SFM tip and to the sample surface, making it most suitable for nanolithography. \cite{Fontaine98} However, for the formation of a water meniscus the tip-sample distance must be significantly reduced compared to scanning operation. The set-point (normalized to the vacuum amplitude of the cantilever, i.e. without tip-sample interaction) in the case of DC oxidation is a highly sensitive parameter. In the DC oxidation (with feedback mechanism on) only a small parameter window exists at which oxidation takes place: For set-points below 8 \% the feedback mechanism gets unstable eventually leading to a tip crash. For set-points above 16 \% the oxidation does not work any more, because the resulting electric field is too low for the formation of a water meniscus.\cite{Garcia98} The low set-point, compared to the 80 \% used in non-contact scanning, is due to the additional electrostatic damping once the voltage is turned on. For a modulated voltage, the average electric field is smaller, allowing higher set-point values without losing the capability of oxidation [see Table~\ref{tab:setpoint}]. In all data (except for Table~\ref{tab:setpoint}) a set-point of 10 \% has been used.

The quantitative discussion about the set-point is specific to the tip, the scanning force microscope model and the sample surface used in the experiment. The fact that for DC and AC oxidation we find an optimal set-point of 10 \% suggests that the decrease of the oscillation amplitude during oxidation is similar in both cases. While for DC oxidation in non-contact mode the oscillation amplitude and the tip-sample distance have been investigated for absolute values,\cite{Garcia98,Calleja99} similar data for the cantilever dynamics of the AC oxidation in non-contact mode have not been reported.

\begin{table}
\caption{\label{tab:setpoint} Cantilever oscillation amplitude in percentage of the vacuum amplitude (without interaction tip-surface). For higher set-points the oxidation failed. Parameters used: $t_{res}$/($t_{res}$+$t_{ox}$)=30 \%, $f$=1 kHz, $V_{ox}$=-16 V ,$V_{res}$=4 V. SFM tip used: Ti-Pt.}
\begin{ruledtabular}
\begin{tabular}{ccc}
			set-point (\%) & height (nm) & aspect ratio\\
			\hline
			10 & $12.9\pm1.0$ & $0.1045\pm0.0085$ \\
			20 & $12.5\pm1.3$ & $0.1052\pm0.0128$ \\
			30 & $12.7\pm0.5$ & $0.1070\pm0.0063$ \\
			40 & $11.8\pm0.6$ & $0.1018\pm0.0088$ \\
\end{tabular}
\end{ruledtabular}
\end{table}

\begin{figure}
\includegraphics[width=0.4\textwidth]{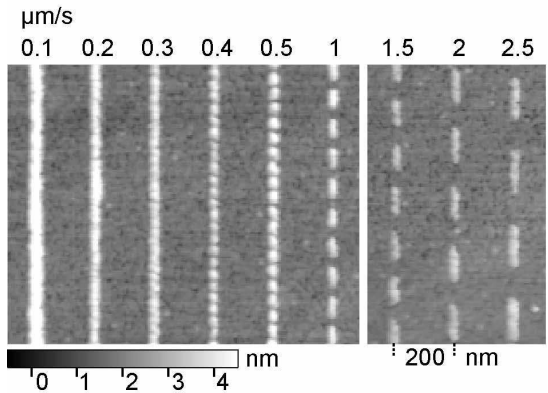}
\caption{\label{fig:velocity} SFM image of oxide lines for different writing speeds at the fixed modulation frequency of $f$=10 Hz. Note that for higher writing speeds the lines get weaker because of the shorter oxidation time.}
\end{figure}

The writing speed competes with the modulation frequency. Fig.~\ref{fig:velocity}, which shows data at a fixed modulation frequency $f$ and increasing writing speed $v$, must be compared to Fig.~\ref{fig:linesfreq} with fixed $v$ but increasing $f$. For small $v$ and large $f$ we are in the limit relevant for the space charge theory, where cycling of reset and oxidation stage becomes relevant and the \textit{individual} dots merge to form a homogeneous oxide line. For large $v$ and small $f$, the cycling gets visible lithographically, the oxidized sections are essentially written with a constant voltage. In all data [except for Fig.~\ref{fig:velocity} and inset Fig.~\ref{fig:frequency}(a)] a writing speed of 0.2 $\mu$m/s has been used.

\subsubsection{\label{sec:reli}Reliability and reproducibility}

\begin{figure}
\includegraphics[width=0.4\textwidth]{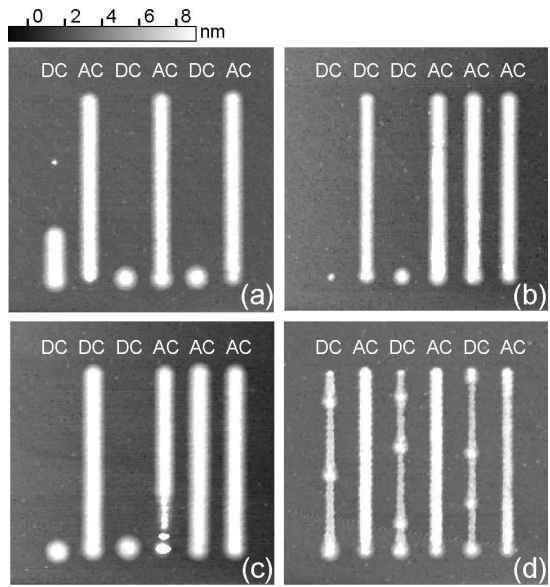}
\caption{\label{fig:reliability} Some characteristic examples for the improved reliability of AC writing. Lines are written from bottom to top and from left to right. Upper figures: DC and AC lines written alternately (a) or successively (b). Lower figures: First AC line with starting difficulty (c), and important inhomogeneity in all three DC lines (d).}
\end{figure}

In a set of 378 1-$\mu$m-long lines, with half of the lines written with AC and half with DC voltage, the success rate of complete and homogeneous lines could be improved up to 95.8 \% for AC writing compared to 78.3 \% for DC oxidation. Four representative examples are shown in Fig.~\ref{fig:reliability}. One possible explanation for the improved reliability is related to the formation of the water meniscus between tip and sample providing the oxyanions for the oxidation process. Turning on the voltage will polarize the water film to form a water bridge. \cite{Garcia99} A threshold voltage is needed to actually start the oxidation, although a lower voltage could be used subsequently for the anodic reaction.\cite{Garcia98} In AC oxidation we have the chance to pick up the water film in each cycle [ Fig.~\ref{fig:reliability}(c) first AC line]. In the DC mode, however, once it failed at the beginning, the strong additional electrostatic damping will keep the tip too far from the sample surface for a water bridge to form. 

Apart from the reliability we find that the reproducibility using modulated voltage as well is better. The standard deviations listed in Table~\ref{tab:aspect} for the oxide height, width and aspect ratio are significantly smaller for AC oxidation. Not only could we enhance the success rate of oxidation, but we find for a given set of parameter a smaller variation in oxide shapes. Several reasons can be listed for the remaining shape fluctuations of successive lines: local differences in the thickness of the water film or native oxide layer, topographic or structural inhomogeneity at the sample surface or abrupt changes in the shape of the SFM tip. All of them will eventually alter the electric field between biased tip and grounded substrate.

\subsection{\label{sec:algas}\normalsize G\scriptsize a\normalsize [A\scriptsize l\normalsize ]A\scriptsize s \normalsize heterostructure: Electronic properties}

In order to investigate the electronic properties of structures fabricated with AC oxidation, we patterned a Al$_{x}$Ga$_{1-x}$As-GaAs heterostructure (Molecular Beam Epitaxy) with a shallow 2DEG 34 nm below the surface (and a remote Si-$\delta$ doping layer with N$_{D}$=1.2$\times 10^{13} $cm$^{-2}$ at half way) and with a mobility $\mu$=30 m$^2$/Vs at 4.2 K. A back gate 1.4 $\mu$m below the 2DEG and a metallic top gate (Ti-Au-film) provide a tunability of the electron density from $N_{2D}$=1.8 up to 3.8 $\times 10^{15}$ m$^{-2}$. The cap layer of 5 nm GaAs is undoped.

There are two important issues for any lithographical technique: accurate projection of the lithographic pattern into the 2DEG and reliable lateral insulating properties. Compared to split gate techniques the confinement potential has steep walls implying a small depletion length and thus an electronic confinement close to the lithographic pattern. \cite{Fuhrer01} Together with a better aspect ratio even more dense nanostructures coupled via multiple terminals come into reach.\cite{Leturcq04}

\begin{figure}
\includegraphics[width=0.4\textwidth]{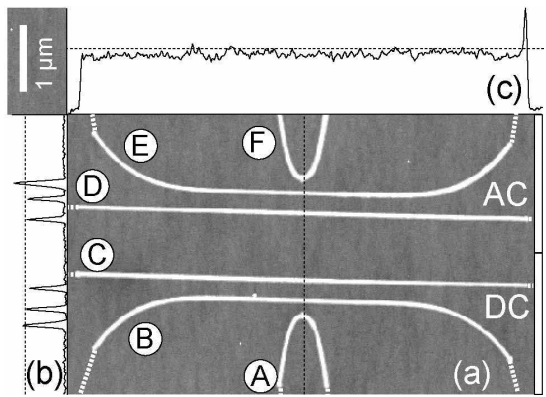}
\caption{\label{fig:structure} (a) SFM image taken before connecting the structure to the outer contact pads (added dashed lines). Two identical structures, quantum wire (width=0.3 $\mu$m, length=3 $\mu$m) and quantum point contact (width=0.21 $\mu$m), have been written with DC voltage (bottom) and AC voltage (top), separated by a 1-$\mu$m-wide channel. (b) Cross section (indicated with a dashed line in the main figure).  (c) Longitudinal section of the AC line D. In both cases the dashed line marks 10 nm in height. Writing parameters: $V_{ox}$=-23 V, $V_{res}$=4 V, $t_{res}$/($t_{res}$+$t_{ox}$)=50 \%, $f$=1 kHz (AC) / $V_{ox}$=-22 V (DC). Set-point 10 \% and $v$=0.2 $\mu$m/s are for both the same. Tip: Ti-Pt}
\end{figure}

In order to investigate the electronic properties and possible differences in DC and AC oxidation we chose to fabricate a (diffusive) quantum wire and a (ballistic) quantum point contact (QPC) (Fig.~\ref{fig:structure}). We were concerned to avoid any additional effect which could mask the difference between DC and AC writing: The two sets of structures (QPC, wire) have therefore been written in the same oxidation process (with a programmable function generator) and close together on the same wafer avoiding most of the fluctuations in ambient humidity and doping concentration. Magnetoresistance and conductance measurements are presented in Sec.~\ref{sec:conf}. First, in Sec.~\ref{sec:break}, the insulating properties of the oxide lines are compared. 

\subsubsection{\label{sec:break}Breakdown voltage of the oxide lines}

\begin{figure}
\includegraphics[width=0.4\textwidth]{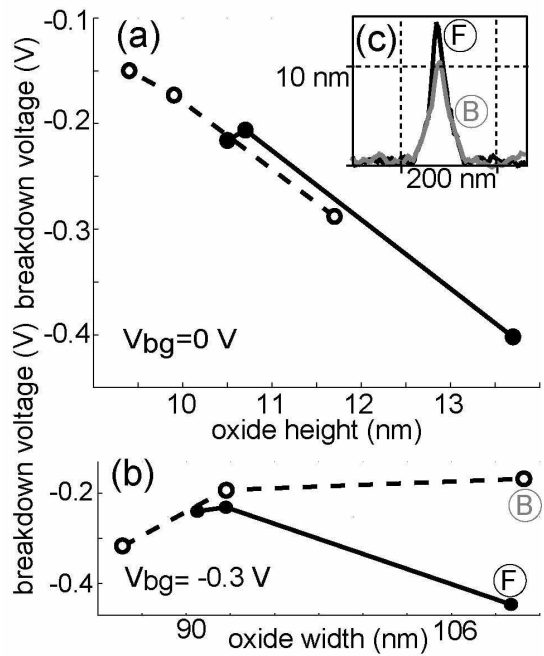}
\caption{\label{fig:breakdown} (a) Breakdown voltages (leakage current of 10 pA) as a function of oxide height (back gate voltage set to 0 V) and (b) breakdown voltages for different oxide widths (back gate voltage set to -0.3 V). Solid lines, $\bullet$: AC writing. Dashed lines, $\circ$: DC writing. The six data points correspond to the six oxide lines of the structure shown in Fig.~\ref{fig:structure}. Measurement done at 4.2 K. (c) Profile of the oxide lines B and F (see Fig.~\ref{fig:structure}): Same base width, but different height in the center of the oxide leading to more than a doubling of the breakdown voltage.}
\end{figure}

In order to evaluate the insulating properties of the depleted regions created by the oxide lines we study the current-voltage curves across the tunnel barrier induced by the oxide lines. Beyond some bias value, called breakdown voltage, the leakage current increases exponentially.
Experimentally we determine the breakdown voltage from the condition that the leakage current reaches 10 pA. 
In Fig.~\ref{fig:breakdown}(a) and (b) the six breakdown voltages for the six oxide lines used in our structure (Fig.~\ref{fig:structure}) are plotted as a function of their oxide heights and widths. We emphasize the fact that the oxide lines connecting the structure to the outer contact pads are higher and thus not limiting.
From Fig.~\ref{fig:breakdown}(a), a quasi linear dependence between oxide height and breakdown voltage can be deduced. Plotting the same points as a function of the base width in Fig.~\ref{fig:breakdown}(b) shows that different values for the breakdown voltage are possible for the same oxide width. 
Comparing the profiles of the oxide lines B (DC) and F (AC) (inset Fig.~\ref{fig:breakdown}), we clearly see that they differ by their central part given by the self-limiting growth.  They are thus good examples for the result shown in Table~\ref{tab:aspect}. An improvement of 26 \% in the aspect ratio, i.e. the same base width but higher central part of the oxide pillar, leads to a giant increase in the breakdown voltage of the oxide line of 186\%.
The reproducibility of the writing process mentioned in Sec.~\ref{sec:reli} and shown in Table~\ref{tab:aspect} turns out to be crucial for the electronic properties. The breakdown voltage and thus the tunability are sensitive to the actual oxide shape. 

\subsubsection{\label{sec:conf}Electronic confinement potential}

At low magnetic field the energy scale of the Landau level separation gets comparable to the level spacing of the electric subbands in the quantum wire leading to a hybridization of magnetic and electric subbands for decreasing magnetic fields. Assuming a parabolic confinement potential, the electronic wire width, the 1D electron density and the potential curvature can be extracted.\cite{Berggren88} Comparing the data (not shown) for several independent cool-downs at 1.7 K and a multitude of in-plane, back and top gate voltages, no qualitative nor quantitative differences between the AC and DC wire could be measured. The steepness of the potential walls does not seem to be altered by the writing scheme. 
Diffusive scattering at rough wire boundaries would lead to an anomalous low-field magnetoresistance peak due to a purely geometrically enhanced backscattering probability.\cite{Thornton89} The lack of such a \textit{wire peak} suggests that the highly specular scattering from the potential wall is not lost for either oxidation technique. Neither lithographic (Fig.~\ref{fig:linesfreq}) nor electronic evidence of discreteness of the oxide line at $f$=1 kHz has been found.

For reduced longitudinal size, inter-level mixing of the transverse modes gets less important. In the limit of purely ballistic constrictions, so-called quantum point contacts (QPC), the population of consecutive electrostatic subbands is reflected in quantized ($2e^2/h$) steps of the conductance as a function of the Fermi level.\cite{Wees88} For a fixed electron density the transition between two quantized plateaus driven by an in-plane gate provides a quantitative measure for the tunability and potential shape in the few mode regime. In Table~\ref{tab:plungergate} lever arms for different gate combinations on the AC and DC QPC are listed. A comparison of the different lever arms leads to the conclusion that the tunability is comparable for all different gating possibilities in all types of structures.

\begin{table}
\caption{\label{tab:plungergate} In-plane gate (IPG) voltage to overcome the quantized energy gap between the third and fourth spin degenerated mode in the QPC for different gate combinations. QPC defined by the wire edge (denoted as \textit{wire}-gate) and v-shaped oxide line (denoted as \textit{qpc}-gate) (Fig.~\ref{fig:structure}). Measurements done at 4.2 K.}
\begin{ruledtabular}
\begin{tabular}{c c c}
			 & $\Delta {V_{IPG}}^{AC}$ (V) & $\Delta {V_{IPG}}^{DC}$ (V) \\
			\hline
			\textit{qpc}-gate & 0.454 & 0.423 \\
			\textit{wire}-gate & 0.191 & 0.199 \\
			\textit{qpc+wire}-gate & 0.147 & 0.158\\
\end{tabular}
\end{ruledtabular}
\end{table}

\section{\label{sec:conc}Conclusion}

In this paper we have shown, how improved control over the oxide growth kinetics can be achieved employing a modulated tip-sample voltage. Our results are compatible with the space charge model proposed by Dagata et al.\cite{Dagata98} We showed enhanced line aspect ratios and also better reliability and improved reproducibility applying modulated voltage on undoped GaAs. These results motivated us to take advantage of this approach for depleting a shallow 2DEG in a Ga[Al]As heterostructure. The electronic stability, the confinement potential and the tunablity were investigated in a wire and a point contact geometry. They are found to be qualitatively and quantitatively similar for oxidation with constant and modulated voltage.

\begin{acknowledgments}
The authors wish to thank J.A. Dagata for pointing out to us the advantages of modulated voltages during the oxidation process. 
Financial support from the Swiss Science Foundation (Schweizerischer Nationalfonds) is gratefully acknowledged. 
\end{acknowledgments}

\end{document}